\documentclass[letterpaper,11pt]{article}

\usepackage{amsthm}
\usepackage{amssymb}
\usepackage{amsmath}
\usepackage{amsfonts}
\usepackage{amstext}

\usepackage[sort,numbers]{natbib}

\usepackage{pxfonts}
\usepackage{eulervm}

\usepackage{fullpage}

\usepackage{color}

\newcommand{\tinyspace}{\mspace{1mu}}

\newcommand*{\bra}[1]{\langle #1|}
\newcommand*{\ket}[1]{|#1\rangle}
\newcommand{\ketbra}[1]{\ket{#1}\bra{#1}}
\newcommand{\braket}[2]{\langle #1|#2\rangle}

\usepackage{dsfont}
\newcommand{\openone}{\mathds{1}}

\newcommand{\tprod}{\otimes}

\newcommand{\norm}[1]{\left\lVert\tinyspace#1\tinyspace\right\rVert}
\newcommand{\tnorm}[1]{\norm{#1}_{\mathrm{tr}}}

\newcommand{\opnorm}[1]{\norm{#1}_{\infty}}

\newcommand{\abs}[1]{\left\lvert\tinyspace #1 \tinyspace\right\rvert}

\newcommand{\dm}[1]{\dim \mathcal{#1}}

\newcommand{\density}[1]{\mathbf{D}(\mathcal{#1})}

\newcommand{\transform}[1]{\mathbf{T}(\mathcal{#1})}
\newcommand{\id}{\openone}
\newcommand{\identity}[1]{\id_{\mathcal{#1}}}

\newcommand{\tidentity}[1]{I_{\mathcal{#1}}}

\newcommand{\rank}{\operatorname{rank}}
\newcommand{\tr}{\operatorname{tr}}
\newcommand{\F}{\operatorname{F}}
\newcommand{\choi}{\operatorname{C}}

\newcommand{\ptr}[1]{\tr_\mathcal{#1}}

\newcommand{\prob}[1]{\textup{\textsc{#1}}}
\newcommand{\class}[1]{\textup{\textsf{#1}}}

\newtheorem{theorem}{Theorem}

\newtheorem{proposition}[theorem]{Proposition}
\newtheorem{corollary}[theorem]{Corollary}
\theoremstyle{definition}

\newtheorem{proto}[theorem]{Protocol}
\newtheorem{problem}[theorem]{Problem}

\usepackage{hyperref}

\title{Testing non-isometry is \class{QMA}-complete}

\author{Bill Rosgen \\
             Centre for Quantum Technologies \\
             National University of Singapore}

\date{1 June, 2010}

\begin{document}

\maketitle

\begin{abstract}
  Determining the worst-case uncertainty added by a quantum circuit is
  shown to be computationally intractable.  This is the problem of
  detecting when a quantum channel implemented as a circuit is close
  to a linear isometry, and it is shown to be complete for the
  complexity class~\class{QMA} of verifiable quantum computation.
  The main idea is to relate the problem of detecting when a channel is
  close to an isometry to the problem of determining how mixed the
  output of the channel can be when the input is a pure state.
\end{abstract}

%=====================================================%

\section{Introduction}

A linear isometry $U \colon \mathcal{H} \to \mathcal{K}$ is a linear map that preserves
the inner product of any two elements, or equivalently satisfies $U^* U = \identity{H}$. 
These transformations are fundamental in quantum
computation: they are exactly the maps that may be realized using
unitary quantum circuits with access to ancillary qubits in a known
pure state---the standard model of quantum computation.  
It is an important problem to
determine when a computation in a non-unitary model, such as
measurement based quantum computing or computation in the presence of
noise, approximately implements some operation in the unitary circuit model.  In this paper
it is shown that this problem is \class{QMA}-complete when the input
computation is modelled as a quantum circuit consisting of the usual unitary gates, plus the ability to discard qubits as well as introduce ancillary qubits.  
The circuit model is not essential: the hardness result also applies
to any model that can efficiently simulate and be simulated by the
mixed-state circuit model.

The complexity class
\class{QMA} is the quantum analogue of \class{NP}: the class
corresponding to classically verifiable computation. This concept
was first considered in~\cite{Knill96}, first formally defined
in~\cite{Kitaev99}, and first studied in~\cite{Watrous00}. \class{QMA}
is the class of all problems that can be verified with bounded error by a polynomial-time
quantum verifier with access to a quantum proof.  This proof is given
by a quantum state on a polynomial number of qubits and may depend on the
input.

The class \class{QMA} has complete (promise) problems:
problems in \class{QMA} that are computationally at least as hard as any other
problem in the class.  This implies that an efficient algorithm for
any of these complete problems can be used to find an efficient
algorithm for any problem in \class{QMA}.
The simplest of these complete problems is the 2-local Hamiltonian problem,
which is informally the quantum version of the circuit satisfiability problem
for unitary circuits with gates of constant size.  A formal description
of this problem, as well as a proof that the 5-local Hamiltonian
problem is \class{QMA}-complete can be found in~\cite{KitaevS+02}.
The improvement of this result to the 2-local case is due to Kempe,
Kitaev, and Regev~\cite{KempeK+06}.  Several other complete problems
for \class{QMA} are known, such as local consistency~\cite{Liu06} (see
also~\cite{LiuC+07, WeiM+10}), some problems related to the minimum
output entropy~\cite{BeigiS07}, testing whether unitary circuits are
close to the identity~\cite{JanzingW+05} (see also~\cite{JiW09}), and
finding the ground states of some physical systems~\cite{SchuchV09, SchuchC+08}.
In the present paper we add a new complete problem to this list: the
problem of determining if a quantum circuit implements an operation
that is close to an isometry.  As discussed in
Section~\ref{scn:channels}, this is equivalent to determining if the channel always
maps pure states to states that are approximately pure.

The remainder of the paper is organized as follows.
Section~\ref{scn:prelim} introduces notation and background.  
Section~\ref{scn:channels} introduces the notion of approximate
isometries and makes formal the problem of
detecting when a channel is an approximate isometry.  The
\class{QMA}-hardness of this problem is proved in
Section~\ref{scn:qma-hardness} and proof of the containment in
\class{QMA}, the most technical portion of the result, 
appears in Section~\ref{scn:qma-containment}.

\section{Preliminaries}\label{scn:prelim}

In this section the notation and background that is used
throughout the paper are presented.  Much of the notation used here is
standard and this is in no way a complete introduction to quantum information.
See~\cite{NielsenC00} for a more detailed treatment of these topics.

All Hilbert spaces considered in this paper are assumed to be
finite-dimensional and are denoted by scripted capital letters $\mathcal{H, K, \ldots}$.
The pure states are the unit vectors in these spaces.
The set of density matrices or mixed states on $\mathcal{H}$ is given by $\density{H}$, and the set of all quantum channels mapping $\density{H}$ to $\density{K}$ is 
$\transform{H,K}$.  The quantum channels are exactly the completely positive
and trace preserving linear maps. 
The identity channel in $\transform{H,H}$ is denoted $\tidentity{H}$, while
$\identity{H}$ is the identity on $\mathcal{H}$.

Given a quantum channel $\Phi \in \transform{H,K}$ we make use of
two representations.  The first of these is the Choi
representation~\cite{Choi75}, which provides a unique representation
of a channel $\Phi \in \transform{H,K}$
as a linear operator on $\mathcal{K \tprod H}$.  
This representation is given by $\choi(\Phi) = (\Phi
\tprod \tidentity{H})(\ketbra{\phi^+})$, where $\ket{\phi^+} = \sum_i
\ket{ii} / \sqrt{d}$ is a maximally entangled state in $\mathcal{H
  \tprod H}$.

The second representation that we use is the representation of a completely
positive map $\Phi$ by a set of Kraus operators: matrices
$A_i$ such that $\Phi(X) = \sum_i A_i X A_i^*$.
This representation is also due to Choi~\cite{Choi75}.
If in addition the map $\Phi$ is trace preserving, then the operators $A_i$
satisfy the property
$\sum_i A_i^* A_i = \identity{}$.
The number of Kraus operators in a
minimal Kraus decomposition is given by the rank of the Choi matrix $\choi(\Phi)$.

In order to measure how close a state is to being pure we use the
operator norm $\opnorm{X}$, which for a linear operator $X$ is
the largest singular value of $X$.  When $X$ is normal, this
is simply the largest eigenvalue (in absolute value) of $X$.
Dual to the operator norm is the trace norm, which for a linear
operator $X$ is given by 
$\tnorm{X} = \tr \sqrt{X^*X}$.
This is exactly the sum of the singular
values of $X$.  When $X$ is a quantum state, this simplifies to the
sum of absolute values of the eigenvalues of $X$, so that $\tnorm{\rho} = 1$ for all
density matrices $\rho$.

One final quantity that we use is the fidelity, which for two
density matrices $\rho, \sigma$ is given by $\F(\rho, \sigma) = \tr
\sqrt{ \sqrt{\rho} \sigma \sqrt{\rho} }$.  While it is not obvious
from this definition, the fidelity is symmetric in the two arguments.
When one of the arguments is a pure state, the fidelity simplifies to
$\F(\rho, \ketbra{\psi}) = \sqrt{ \bra\psi \rho \ket\psi }$.
An important relationship between the trace norm and the fidelity
is
\begin{equation}\label{eqn:fid-tnorm-onepure} 
  2 - 2 \F(\rho, \ketbra{\psi})^2 \leq \tnorm{\rho - \ketbra\psi}
\end{equation}
that we use to relate different notions of the purity of a quantum state.  This inequality can
be found in~\cite[Chapter 9]{NielsenC00}.

We require one final piece of background.  In order for a quantum
channel to be given as input to a computational problem we need a
representation of the channel.  Using either the Choi matrix or Kraus
operators produces a representation that, in the case of channels
implementing efficient quantum algorithms, is exponentially larger
than the size of a circuit representation.  These channels have circuit representations 
that are logarithmic in the
Hilbert space dimension.  For this reason, we use a circuit
representation of quantum channels.
Such a representation is provided by the mixed-state circuit model of
Aharonov et al.~\cite{AharonovK+98}, which is simply the usual model
of unitary quantum circuits with two additional gates.  These gates
are the gate that introduces ancillary qubits in the $\ket 0$ state
and the gate the traces out (i.e.\ discards) a qubit.  This circuit
model can be used to represent any quantum channel, which makes it
ideal for the problem that we consider.

\section{Isometries and rank non-increasing channels}\label{scn:channels}

One important property of the linear isometries is that they do not
increase rank.  This is essential to the \class{QMA} protocol
in Section~\ref{scn:qma-containment}, which is able to detect exactly
those channels that are rank-increasing.  More formally, a channel
$\Phi$ is \emph{rank non-increasing} if for all states $\rho$ the
output of $\Phi$ satisfies $\rank(\rho) \geq \rank( \Phi(\rho) )$.
Unfortunately, this property does not characterize the isometries.  
Consider the channel $\Phi(\rho) = \ket 0 \bra 0$ that
discards the input state and returns a fixed pure state.  This channel
is not an isometry but it is also rank non-increasing.

This property can be used to characterize the isometries if we make a
small adjustment.  The channels that are rank non-increasing when
adjoined to an auxiliary space of arbitrary dimension are exactly the
isometries.  We call a channel $\Phi \in \transform{H,K}$ 
\emph{completely} rank non-increasing
if for any $\mathcal{F}$ the channel
  $\Phi \tprod \tidentity{F}$ is rank non-increasing, i.e.\ if 
   $\rank \left[ (\Phi \tprod \tidentity{F})(\rho) \right] \leq \rank ( \rho )$ for all $\rho$.
The channel $\Phi(\rho) = \ket 0 \bra 0$ is not completely rank
non-increasing: consider applying it to half of a maximally entangled state
 $ (\Phi \tprod \tidentity{H})(\ket{ \phi^+ } \bra{ \phi^+ })
    = \ket 0 \bra 0 \tprod \identity{H} / \dm{H} $.
As in the case of complete positivity, we need only to verify this property
on an auxiliary space of the same dimension as the input space.  It is also easy to see that
this property characterizes the linear isometries.
\begin{proposition}\label{prop:equiv-noerror}
  The following are equivalent for a channel $\Phi \in \transform{H,K}$:
  \begin{enumerate}
    \item $\Phi(\rho) = U \rho U^*$ for some linear isometry $U$ from
      $\mathcal{H}$ to $\mathcal{K}$,
      \label{enum:crni-1}
    \item $\Phi$ is completely rank non-increasing,
      \label{enum:crni-2}
    \item $\Phi \tprod \tidentity{H}$ is rank non-increasing.
      \label{enum:crni-3}
  \end{enumerate}
\end{proposition}
\begin{proof}
  The first two implications are immediate.
  To prove that $\eqref{enum:crni-3}
  \Rightarrow \eqref{enum:crni-1}$, let $\Phi \tprod
  \tidentity{H}$ be rank non-increasing.  This implies that $\rank (\choi(\Phi)) = 1$.  Recalling that the number of Kraus operators in a
  minimal decomposition is $\rank (\choi(\Phi))$, it follows that $\Phi$ can be
  expressed as $\Phi(\rho) = A \rho A^*$.
  The condition that $\Phi$ is trace preserving implies that the
  operator $A$ satisfies $A^* A = \identity{H}$. \qed
\end{proof}

This characterization guides the remainder of
the paper.  Detecting when the channel $\Phi \tprod \tidentity{H}$
increases rank provides an operational method to determine when a
channel is an isometry.

\subsection{Approximately pure states}\label{scn:state-purity}

In order to show that non-isometry detection is \class{QMA}-complete
we need to consider an approximate version of the problem.
This is because a protocol for a \class{QMA} language is
permitted to fail with small probability.  The definition of
approximate isometries used here is closely related to the notion of
approximately pure states.  Several equivalent notions of the purity
of a density matrix are considered in this section.

Perhaps the most well-known notion of how close a mixed state $\rho$ is to
being pure is the \emph{purity} of $\rho$,
given by $\tr(\rho^2)$.  A similar measure is given by $\opnorm{\rho}$, 
the largest
eigenvalue of $\rho$.
It is not hard to see that these quantities are related.  If $\rho =
\sum_i \lambda_i \ket{\lambda_i}\bra{\lambda_i}$ is the spectral
decomposition of $\rho$, with the eigenvalues $\lambda_i$ in
decreasing order, then
$\tr{\rho^2} = \sum_i \lambda_i^2 
  \geq \lambda_1^2 
  = \opnorm{\rho}^2$.
In the other direction, since the purity is convex, it is maximized
for $1 / \lambda_1$ eigenvalues each of value $\lambda_1$, i.e.\
$\tr{\rho^2} 
  = \sum_i \lambda_i^2 
  \leq \lambda_1^2 / \lambda_1
  = \opnorm{\rho}$.
Taken together, these two inequalities show that
\begin{equation}\label{eqn:purity-largest-eval}
  \opnorm{\rho}^2 \leq \tr(\rho^2) \leq \opnorm{\rho}.
\end{equation}
These quantities are also related to the more familiar trace
distance on quantum states.
\begin{proposition}\label{prop:opnorm-tnorm}
  Let $\rho \in \density{H}$ and let $\varepsilon > 0$.  There exists a
  pure state $\ket \psi \in \mathcal{H}$ such that $\tnorm{\rho -
    \ket\psi \bra\psi } \leq \varepsilon$ if and only if
  $\opnorm{\rho} \geq 1 - \varepsilon / 2$.
\end{proposition}
\begin{proof}
  Let $\rho$ have spectral decomposition given by $\rho = \sum_i
  \lambda_i \ket{\lambda_i}\bra{\lambda_i}$, with
  $\lambda_1 \geq \lambda_2 \geq \ldots \geq \lambda_d$.
  If $\opnorm{\rho} = \lambda_1 \geq 1 - \varepsilon/2$, then
  \begin{align*}
    \tnorm{ \rho - \ket{\lambda_1}\bra{\lambda_1} }
    = (1 - \lambda_1) + \sum\nolimits_{i=2}^d \lambda_i
    = 2 (1 - \lambda_1)
    \leq 2 (\varepsilon / 2)
    = \varepsilon.
  \end{align*}
  On the other hand, if $\ket\psi \in \mathcal{H}$ is a state such
  that $\tnorm{\rho - \ket\psi\bra\psi} \leq \varepsilon$, then
  by Equation~\eqref{eqn:fid-tnorm-onepure}
  \begin{align}
    \varepsilon
    \geq \tnorm{\rho - \ket\psi\bra\psi}
    \geq 2 - 2 \F(\rho, \ket\psi\bra\psi)^2
    = 2 - 2 \bra{\psi} \rho \ket{\psi}
    = 2 - 2 \sum\nolimits_i \lambda_i \abs{ \braket{\psi}{\lambda_i} }^2.
    \label{eqn:prop-tnorm-purity-eps-to-convex}
  \end{align}
  The final quantity is a convex combination of the $\lambda_i$, with
  weights determined by the state $\ket\psi$.  This is maximized
  when $\ket\psi = \ket{\lambda_1}$, since $\lambda_1$ is the largest
  eigenvalue of $\rho$.  Combining this with
  Equation~\eqref{eqn:prop-tnorm-purity-eps-to-convex} we have
  $ \varepsilon \geq 2 - 2 \lambda_1 = 2 - 2 \opnorm{\rho} $,
  which implies that $\opnorm{\rho} \geq 1 - \varepsilon / 2$. \qed
\end{proof}

Given these notions of purity, we will call a state \emph{$\varepsilon$-pure}
if $\opnorm{\rho} \geq 1 - \varepsilon$.  By the previous results the
purity of such a state satisfies $\tr(\rho^2) \geq (1 - \varepsilon)^2
\geq 1 - 2 \varepsilon$, and there is a pure state $\ket\psi$ such
that $\tnorm{\rho - \ket\psi \bra\psi} \leq 2 \varepsilon$.  For the
results of this paper, any of these three measures suffices, as they
are equivalent up to polynomial factors in $\varepsilon$.

\subsection{Approximate isometries}

The focus of this paper is to show that detecting when a channel is
far from an isometry is computationally difficult.  To do this we
need to define the class of channels that are the
approximate isometries.
Isometries always map pure states to pure states, even in the presence of a
reference system.  Proposition~\ref{prop:equiv-noerror} 
shows that this condition characterizes the isometries.
Weakening this requirement, we call a channel an
\emph{$\varepsilon$-isometry} if it maps pure states (over the input space and a
reference system) to states that are $\varepsilon$-pure, for some
$\varepsilon > 0$.

More formally a channel $\Phi \in \transform{H,K}$ is an
$\varepsilon$-isometry if for any pure state $\ket\psi \in \mathcal{H
  \tprod H}$ the output of $\Phi \tprod \tidentity{H}$ satisfies
$\opnorm{(\Phi \tprod \tidentity{H})(\ket\psi\bra\psi)} \geq 1 -
\varepsilon$, i.e.\ when applied to
part of any pure state the output state is close to pure.
This implies that $\Phi \tprod \tidentity{H}$ does not reduce the
operator norm of any input by more than a factor of $1 - \varepsilon$.
We use this to define the
computational problem that is the main focus of the paper.
\begin{problem}[\prob{Non-isometry}]
  For $0 \leq \varepsilon < 1/2$ and a channel $\Phi \in
  \transform{H,K}$, given as a mixed-state quantum circuit, the promise
  problem is to decide between:
  \begin{description}
    \item[Yes:] There exists a pure state $\ket\psi \in \mathcal{H}$ such that $\opnorm{(\Phi
        \tprod \tidentity{H})(\ketbra{\psi})} \leq \varepsilon$,
    \item[No:] For all pure states $\ket\psi \in \mathcal{H}$, $\opnorm{(\Phi
        \tprod \tidentity{H})(\ketbra{\psi})} \geq 1 - \varepsilon$.
  \end{description}
  When the value of $\varepsilon$ is significant, we will refer to this
  problem as $\prob{Non-isometry}_\varepsilon$.
\end{problem}

Using the equivalence results of
Equation~\eqref{eqn:purity-largest-eval} and
Proposition~\ref{prop:opnorm-tnorm}, this problem may be equivalently
defined in terms of either the purity or the trace distance to the
closest pure state, up to a small increase in $\varepsilon$.
The case of the minimum output purity of a channel has been studied in
a different context by Zanardi and Lidar~\cite{ZanardiL04}, though
they focus on finding the minimum purity of a channel over a subspace
of the inputs.  The problem we consider here is equivalent to
evaluating the channel purity of $\Phi \tprod
\tidentity{H}$ over the whole input space.

The difficulty of the \prob{Non-isometry} problem does not change if
the dimension of the ancillary system is permitted to be larger than the size of the
input system, so long as the number of qubits needed to represent the
ancillary system is polynomial in the number of input qubits.

The notion of approximate isometry that we consider here is \emph{not}
equivalent to the channel being completely rank non-increasing on
average.  This property is modelled
by the distance between the Choi matrix of a channel and a pure state.
While it
is true that the Choi matrix is pure if and only if the channel is an
isometry, it is close to pure in the trace distance when the channel is close to an
isometry \emph{on average}.  In this paper we consider the worst-case,
i.e.\ we consider a channel to be close to an isometry if and only if
the output of $\Phi \tprod \tidentity{H}$ is close to pure for
\emph{any} pure state input.  
A simplification of the protocol presented in
Section~\ref{scn:qma-containment} yields a polynomial-time quantum
algorithm for the problem of determining how close the Choi matrix of
a channel is to a pure state.  This is because $\choi(\Phi)$ can be generated efficiently, and given two copies the swap test can be used to test the purity of a quantum state as shown in~\cite{EkertA+02}.

\section{QMA hardness}\label{scn:qma-hardness}

In order to prove the hardness of \prob{Non-isometry} we modify an
arbitrary \class{QMA} protocol to obtain a circuit that
can output a mixed state exactly when the verifier would have accepted
in the original protocol.  This yields a circuit that is far from an
isometry if and only if there is a witness that causes the verifier in
the original protocol to accept.  Deciding whether or not there is
such a witness is \class{QMA}-hard, by the definition of the
complexity class.
More formally, a language $L$ is in \class{QMA} if there is a
quantum polynomial-time verifier $V$ such that
\begin{enumerate}
  \item if $x \in L$, then there exists a witness $\rho$ such that
    $\Pr[ \text{$V$ accepts $\rho$} ] \geq 1 - \varepsilon$,
  \item if $x \not\in L$, then for any state $\rho$, 
    $\Pr[ \text{$V$ accepts $\rho$} ] \leq \varepsilon$,
\end{enumerate}
The exact value of the error parameter $\varepsilon$ is not
significant: any $\varepsilon < 1/2$ that is at least an inverse
polynomial in the input size suffices~\cite{KitaevS+02, MarriottW05}.

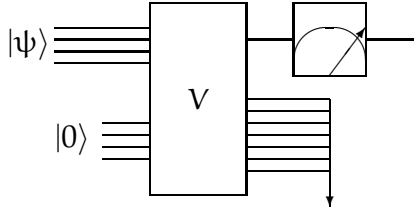
\begin{figure}
   \begin{center}
\setlength{\unitlength}{3947sp}%
\begingroup\makeatletter\ifx\SetFigFont\undefined%
\gdef\SetFigFont#1#2#3#4#5{%
  \reset@font\fontsize{#1}{#2pt}%
  \fontfamily{#3}\fontseries{#4}\fontshape{#5}%
  \selectfont}%
\fi\endgroup%
\begin{picture}(2577,1299)(-764,-1423)
\put(-749,-436){\makebox(0,0)[lb]{\smash{{\SetFigFont{12}{14.4}{\rmdefault}{\mddefault}{\updefault}{\color[rgb]{0,0,0}$\ket \psi$}%
}}}}
\thinlines
{\color[rgb]{0,0,0}\put(751,-1036){\line( 1, 0){525}}
}%
{\color[rgb]{0,0,0}\put(751,-1111){\line( 1, 0){525}}
}%
{\color[rgb]{0,0,0}\put(751,-1186){\line( 1, 0){525}}
}%
{\color[rgb]{0,0,0}\put(751,-886){\line( 1, 0){525}}
}%
{\color[rgb]{0,0,0}\put(751,-811){\line( 1, 0){525}}
}%
{\color[rgb]{0,0,0}\put(751,-736){\line( 1, 0){525}}
}%
{\color[rgb]{0,0,0}\put(1276,-511){\oval(450,450)[tr]}
\put(1276,-511){\oval(450,450)[tl]}
}%
{\color[rgb]{0,0,0}\put(151,-1336){\framebox(600,1200){$V$}}
}%
{\color[rgb]{0,0,0}\put(-449,-286){\line( 1, 0){600}}
}%
{\color[rgb]{0,0,0}\put(-449,-361){\line( 1, 0){600}}
}%
{\color[rgb]{0,0,0}\put(-449,-436){\line( 1, 0){600}}
}%
{\color[rgb]{0,0,0}\put(-449,-511){\line( 1, 0){600}}
}%
{\color[rgb]{0,0,0}\put(-149,-961){\line( 1, 0){300}}
}%
{\color[rgb]{0,0,0}\put(-149,-1036){\line( 1, 0){300}}
}%
{\color[rgb]{0,0,0}\put(-149,-1111){\line( 1, 0){300}}
}%
{\color[rgb]{0,0,0}\put(-149,-886){\line( 1, 0){300}}
}%
{\color[rgb]{0,0,0}\put(1276,-586){\vector( 3, 4){225}}
}%
{\color[rgb]{0,0,0}\put(1051,-586){\framebox(450,450){}}
}%
{\color[rgb]{0,0,0}\put(1051,-361){\line(-1, 0){300}}
}%
{\color[rgb]{0,0,0}\put(1276,-736){\vector( 0,-1){675}}
}%
{\color[rgb]{0,0,0}\put(1501,-361){\line( 1, 0){300}}
}%
\put(-449,-1036){\makebox(0,0)[lb]{\smash{{\SetFigFont{12}{14.4}{\rmdefault}{\mddefault}{\updefault}{\color[rgb]{0,0,0}$\ket 0$}%
}}}}
{\color[rgb]{0,0,0}\put(751,-961){\line( 1, 0){525}}
}%
\end{picture}%

   \end{center}
   \caption{Verifier's circuit in a \class{QMA} protocol.
     The verifier accepts the witness state $\ket\psi$ if and only if
     the measurement in the computational basis results in the $\ket 1$ state.}
   \label{fig:qma}
 \end{figure}

Let $L$ be an arbitrary language in \class{QMA}, and let $x$ be
an arbitrary input string.  The goal is to embed the \class{QMA}-hard
problem of deciding if $x \in L$ into the problem of testing whether
a mixed-state quantum circuit is close to an isometry.
Let $V$ be the isometry
representing the algorithm of the verifier in a \class{QMA}
protocol for $L$ on input $x$.  We may ``hard-code'' the
input string $x$ into $V$ because the circuit needs
only to be efficiently generated from $x$.  
The algorithm implemented by the verifier 
is shown in Figure~\ref{fig:qma}.  
The verifier first receives a witness state $\ket\psi$, applies the
isometry $V$, and then makes a measurement on one of the qubits, the
result of which determines whether or not the verifier accepts.  
Any qubits not measured are traced out.

For concreteness, let $V$ act on the input spaces $\mathcal{W}$ and
$\mathcal{A}$, which hold the witness state and the $\ket 0$ state of
the ancilla respectively.
Let $\mathcal{M}$ be the space
corresponding to the measured output qubit in the protocol and let
$\mathcal{G}$ represent the `garbage' qubits that are traced out at
the end of the protocol.
The probability that verifier accepts the witness state $\ket\psi \in \mathcal{W}$ is
\begin{equation}\label{eqn:qma-accept}
  \Pr [ \text{$V$ accepts $\ket\psi$} ] 
  = \bra 1 \ptr{G} \left[ V (\ket\psi\bra\psi \tprod \ket 0 \bra 0) V^* \right] \ket 1.
\end{equation}
Deciding if there
is some $\ket\psi$ such that this expectation is close to one is complete for \class{QMA}.

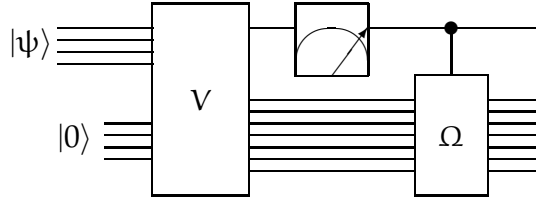
\begin{figure}[t]
  \begin{center}
\setlength{\unitlength}{3947sp}%
\begingroup\makeatletter\ifx\SetFigFont\undefined%
\gdef\SetFigFont#1#2#3#4#5{%
  \reset@font\fontsize{#1}{#2pt}%
  \fontfamily{#3}\fontseries{#4}\fontshape{#5}%
  \selectfont}%
\fi\endgroup%
\begin{picture}(3327,1224)(-764,-1348)
\put(-749,-436){\makebox(0,0)[lb]{\smash{{\SetFigFont{12}{14.4}{\rmdefault}{\mddefault}{\updefault}{\color[rgb]{0,0,0}$\ket \psi$}%
}}}}
\thinlines
{\color[rgb]{0,0,0}\put(1276,-586){\vector( 3, 4){225}}
}%
{\color[rgb]{0,0,0}\put(1051,-586){\framebox(450,450){}}
}%
{\color[rgb]{0,0,0}\put(751,-961){\line( 1, 0){1050}}
}%
{\color[rgb]{0,0,0}\put(751,-1036){\line( 1, 0){1050}}
}%
{\color[rgb]{0,0,0}\put(751,-1111){\line( 1, 0){1050}}
}%
{\color[rgb]{0,0,0}\put(751,-1186){\line( 1, 0){1050}}
}%
{\color[rgb]{0,0,0}\put(751,-886){\line( 1, 0){1050}}
}%
{\color[rgb]{0,0,0}\put(751,-811){\line( 1, 0){1050}}
}%
{\color[rgb]{0,0,0}\put(751,-736){\line( 1, 0){1050}}
}%
{\color[rgb]{0,0,0}\put(2026,-286){\circle*{76}}
}%
{\color[rgb]{0,0,0}\put(2251,-961){\line( 1, 0){300}}
}%
{\color[rgb]{0,0,0}\put(2251,-1036){\line( 1, 0){300}}
}%
{\color[rgb]{0,0,0}\put(2251,-1111){\line( 1, 0){300}}
}%
{\color[rgb]{0,0,0}\put(2251,-1186){\line( 1, 0){300}}
}%
{\color[rgb]{0,0,0}\put(2251,-886){\line( 1, 0){300}}
}%
{\color[rgb]{0,0,0}\put(2251,-811){\line( 1, 0){300}}
}%
{\color[rgb]{0,0,0}\put(2251,-736){\line( 1, 0){300}}
}%
{\color[rgb]{0,0,0}\put(151,-1336){\framebox(600,1200){$V$}}
}%
{\color[rgb]{0,0,0}\put(-449,-286){\line( 1, 0){600}}
}%
{\color[rgb]{0,0,0}\put(-449,-361){\line( 1, 0){600}}
}%
{\color[rgb]{0,0,0}\put(-449,-436){\line( 1, 0){600}}
}%
{\color[rgb]{0,0,0}\put(-449,-511){\line( 1, 0){600}}
}%
{\color[rgb]{0,0,0}\put(-149,-961){\line( 1, 0){300}}
}%
{\color[rgb]{0,0,0}\put(-149,-1036){\line( 1, 0){300}}
}%
{\color[rgb]{0,0,0}\put(-149,-1111){\line( 1, 0){300}}
}%
{\color[rgb]{0,0,0}\put(-149,-886){\line( 1, 0){300}}
}%
{\color[rgb]{0,0,0}\put(751,-286){\line( 1, 0){300}}
}%
{\color[rgb]{0,0,0}\put(1501,-286){\line( 1, 0){1050}}
}%
{\color[rgb]{0,0,0}\put(1801,-1336){\framebox(450,750){$\Omega$}}
}%
{\color[rgb]{0,0,0}\put(2026,-286){\line( 0,-1){300}}
}%
\put(-449,-1036){\makebox(0,0)[lb]{\smash{{\SetFigFont{12}{14.4}{\rmdefault}{\mddefault}{\updefault}{\color[rgb]{0,0,0}$\ket 0$}%
}}}}
{\color[rgb]{0,0,0}\put(1276,-511){\oval(450,450)[tr]}
\put(1276,-511){\oval(450,450)[tl]}
}%
\end{picture}%

  \end{center}
  \caption{Constructed instance of \prob{Non-isometry}.  The output
    state is mixed by the completely depolarizing channel $\Omega$
    only if the state $\ket\psi$ is a valid witness to the original
    \class{QMA} protocol. }
  \label{fig:hardness}
\end{figure}

From Figure~\ref{fig:qma} it is simple to construct a circuit
that produces highly mixed output exactly when there exists such a
$\ket\psi$.  The idea is add a controlled application of the completely
depolarizing channel $\Omega$ on the space $\mathcal{G}$, instead of
tracing it out.  The resulting circuit is shown in Figure~\ref{fig:hardness}.
In the case that the verifier accepts with negligible probability for
every input state $\ket\psi$, then both the measurement and the
controlled depolarizing channel have little effect, leaving the state
of the system close to a pure state.  If, on the other hand, there is
a state on which the verifier accepts with high probability, then on
this input the circuit in Figure~\ref{fig:hardness} produces a highly
mixed state.  Formalizing this notion proves that \prob{Non-isometry}
is \class{QMA}-hard.

\begin{theorem}\label{thm:hardness}
  Let $\varepsilon > 0$ be a constant, and let $p$ be the maximum
  acceptance probability of the protocol $V$.  Let $\Phi
  \in \transform{W,M \tprod G}$ be the circuit in
  Figure~\ref{fig:hardness}.  Then if $\dm{R} = \dm{W}$
  \begin{align*}
    p \leq \varepsilon & \implies 
      \min_{\ket\psi} \opnorm{(\Phi \tprod \tidentity{R})(\ket\psi\bra\psi)} 
      \geq 1 - \varepsilon, \\
    p \geq 1 - \varepsilon & \implies 
      \min_{\ket\psi} \opnorm{(\Phi \tprod \tidentity{R})(\ket\psi\bra\psi)} 
      \leq \varepsilon.
  \end{align*}
\end{theorem}
\begin{proof}  
  Notice that we may assume that the output dimension of $\Phi$ is
  $\dm{M \tprod G} = 2 d > 2 / \varepsilon$ by padding the circuit for
  $V$ with $\log 1/\varepsilon$ unused ancillary qubits, if necessary.
  
  As a first step, we evaluate the output state of the channel $\Phi
  \tprod \tidentity{R}$.
  Applied to a pure state $\ket\psi \in \mathcal{W \tprod R}$ this
  channel first adds the ancillary $\ket 0$ qubits in the space
  $\mathcal{A}$ and then applies the isometry $V$ from the
  \class{QMA} protocol.  This results in the pure state
   $ \ket \phi
    = (V \tprod \identity{R}) 
    (\ket \psi \tprod \ket 0)$.
  We may decompose this state in terms of the qubit in the space
  $\mathcal{M}$, obtaining for some $0 \leq p \leq 1$
  \begin{equation*}
    \ket{\phi}
    = \sqrt{1-p} \ket 0 \tprod \ket{\phi_0}
    + \sqrt{p} \ket 1 \tprod \ket{\phi_1}.
  \end{equation*}
  The value of $p$ is exactly the probability that the measurement
  result is $\ket 1$, i.e.\ the probability that the verifier will
  accept the input state $\ptr{R} \ket\psi \bra \psi$ in the
  original protocol.
  Using this, the state
  after the measurement and the controlled depolarizing channel on $\mathcal{G}$ is
  \begin{equation}\label{eqn:hardness-output}
    (1-p) \ketbra 0 \tprod \ketbra{\phi_0} 
    + (p/d) \ketbra 1 \tprod \identity{G} \tprod \rho,
  \end{equation}
  where $\rho$ is the residual state on $\mathcal{R}$ after this
  channel has been applied ($\rho = \ptr{G} \ketbra{\phi_1}$, but
  this will not be important).  Evaluating the largest eigenvalue of
  this state we find that
  \begin{equation}\label{eqn:hardness-max}
    \opnorm{(\Phi \tprod \tidentity{R})(\ketbra \phi)}
    = \max \{ 1 - p, \frac{p}{d} \opnorm{\rho} \}.
  \end{equation}
  
  We analyze the maximum in Equation~\eqref{eqn:hardness-max} in two
  cases.  The first of these cases is when there is no input
  the verifier accepts with probability larger than $\varepsilon$.  In this case the
  output of the channel $\Phi \tprod \tidentity{R}$
  is given by Equation~\eqref{eqn:hardness-output} where $p
  \leq \varepsilon$.  Here Equation~\eqref{eqn:hardness-max} shows that the
  output has an eigenvalue of magnitude at least
  $\min_{\ket\mu} \opnorm{(\Phi \tprod \tidentity{R})(\ketbra \mu)}
    \geq 1 - p \geq 1 - \varepsilon$.
  
  The second case is when there exists a state
  $\ket\psi$ that
  verifier to accepts with probability at least $1 - \varepsilon$.
  In this case we take the input state to $\Phi \tprod
  \tidentity{R}$ to be $\ket\gamma = \ket\psi \tprod \ket 0$, i.e.\ we set the
  reference system to be any pure state that is not entangled with
  the rest of the input.  The output is given by
  Equation~\eqref{eqn:hardness-output} with $p \geq 1 -
  \varepsilon$ and $\rho = \ketbra 0$.
  Equation~\eqref{eqn:hardness-max} yields
  \begin{equation*}
    \min_{\ket\mu} \opnorm{(\Phi \tprod \tidentity{R})(\ketbra \mu)}
    \leq \opnorm{(\Phi \tprod \tidentity{R})(\ketbra \gamma)} 
    = \max \left\{ 1 - p,  \frac{p}{d} \opnorm{\rho} \right\} 
    \leq \max \left\{ \varepsilon,  \frac{1}{d} \right\} = \varepsilon,
  \end{equation*}
  as we have taken $1 / d < \varepsilon$ (by adding $O(\log
  1/\varepsilon)$ unused ancillary qubits if necessary). \qed
\end{proof}

This theorem shows that determining how far the output $\Phi \tprod
\tidentity{R}$ is from a pure state is as computationally difficult as
determining whether or not the verifier can be made to accept with
high probability in a \class{QMA} protocol.  Since  the
construction of the circuit shown in Figure~\ref{fig:hardness} can be
performed efficiently, this implies the hardness of this problem.

\begin{corollary}\label{cor:noniso-hardness}
  For any constant $0 \leq \varepsilon < 1/2$, \prob{Non-isometry} is \class{QMA}-hard.
\end{corollary}

\noindent Using the equivalences between notions of purity in of
Section~\ref{scn:state-purity}, this also implies that evaluating the
purity of a quantum channel, as defined by Zanardi and
Lidar~\cite{ZanardiL04} is \class{QMA}-hard.

\section{QMA protocol}\label{scn:qma-containment}

In order to show that \prob{Non-isometry} is \class{QMA}-complete, it
remains only to construct a \class{QMA} protocol for the problem.  The
key idea behind this protocol is that when two copies of a channel
$\Phi$ are applied in parallel to the input state $\ket\psi
\tprod \ket\psi$ the output lies in the antisymmetric subspace if and
only if $\Phi(\ketbra \psi)$ is a mixed state.  This
provides a probabilistic test that can detect when a channel is far
from an isometry.

Unfortunately, in a \class{QMA} protocol the verifier cannot assume
the witness is given by two non-entangled pure states.  It suffices,
however, for the verifier to require that the input state lies in the
symmetric subspace of the input space $(\mathcal{H \tprod R})^{\tprod
  2}$.  To show that the channel is not an isometry in \class{QMA},
the prover can provide a symmetric state that a parallel application
of the channel maps into the antisymmetric space of the output space
$(\mathcal{K \tprod R})^{\tprod 2}$.

The verifier in such a protocol needs a test to determine when a state
is symmetric or antisymmetric.  Such a test is provided by the swap
test, which was introduced in the context of communication complexity
in~\cite{BuhrmanC+01}, though we make use of it to test purity using
an idea from~\cite{EkertA+02}.

The swap test can be
characterized as the projection onto the symmetric
and antisymmetric subspaces of a bipartite space.  If $W$ is the swap operation on a
space $\mathcal{H \tprod H}$, then the symmetric measurement outcome of
the swap test corresponds to the the projector
$(\identity{H \tprod H} + W)/2$, and the projector
$(\identity{H \tprod H} - W)/2$ corresponds to the antisymmetric
outcome.

The main idea behind the protocol for \prob{Non-isometry} is that the
swap test can be used to measure the purity of a state.  As
observed in~\cite{EkertA+02}, when applied two to copies of a state $\rho = \sum_i \lambda_i \ketbra{\psi_i}$ the swap test returns the antisymmetric outcome with probability
\begin{align}
  \frac{1}{2} \tr( (\identity{} - W)(\rho \tprod \rho) )
%  &= \frac{1}{2} - \frac{1}{2} \sum_{ij} \lambda_i \lambda_j  
%      \tr\left[ W (\ketbra{\psi_i} \tprod \ketbra{\psi_j}) \right] \nonumber \\
  &= \frac{1}{2} - \frac{1}{2} \sum\nolimits_{i} \lambda_i^2
    =  \frac{1}{2} - \frac{1}{2} \tr(\rho^2). \label{eqn:swaptest-purity}
\end{align}
This implies that the swap test on two copies of a state can be used
to test purity and, by extension, when a channel is far from an
isometry.

A straightforward protocol for $\prob{Non-isometry}$ on a channel
$\Phi$ is then to receive a witness state $\ket\psi \tprod
\ket\psi$, apply the channel to obtain $[(\Phi \tprod
\tidentity{})(\ketbra\psi)]^{\tprod 2}$, and finally apply the swap
test.  The result is the antisymmetric outcome 
with high probability only when the
state $(\Phi \tprod \tidentity{})(\ketbra\psi)$ is highly mixed.
Such a protocol detects the channels that are far from
isometries.

Unfortunately, the verifier in a \class{QMA} protocol cannot assume
that the witness state is of the form $\ket\psi \tprod \ket\psi$.  The
verifier \emph{can} check that he has received some state in the
symmetric subspace and then use the fact that this subspace is closed
under the parallel application of a rank non-increasing channel.  The
verifier in the following protocol uses the swap test to both check
the symmetry of the input state and the antisymmetry of the output
state.

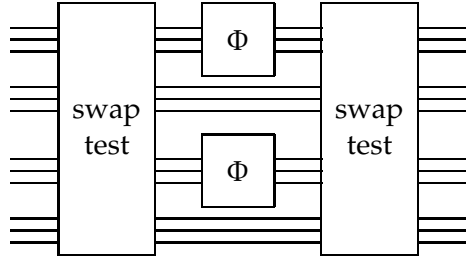
\begin{figure}
  \begin{center}
\setlength{\unitlength}{3947sp}%
\begingroup\makeatletter\ifx\SetFigFont\undefined%
\gdef\SetFigFont#1#2#3#4#5{%
  \reset@font\fontsize{#1}{#2pt}%
  \fontfamily{#3}\fontseries{#4}\fontshape{#5}%
  \selectfont}%
\fi\endgroup%
\begin{picture}(2874,1599)(1339,-1873)
\thinlines
{\color[rgb]{0,0,0}\put(2251,-436){\line( 1, 0){300}}
}%
{\color[rgb]{0,0,0}\put(2251,-586){\line( 1, 0){300}}
}%
{\color[rgb]{0,0,0}\put(1351,-511){\line( 1, 0){300}}
}%
{\color[rgb]{0,0,0}\put(1351,-436){\line( 1, 0){300}}
}%
{\color[rgb]{0,0,0}\put(1351,-586){\line( 1, 0){300}}
}%
{\color[rgb]{0,0,0}\put(2251,-811){\line( 1, 0){1050}}
}%
{\color[rgb]{0,0,0}\put(2251,-886){\line( 1, 0){1050}}
}%
{\color[rgb]{0,0,0}\put(2251,-961){\line( 1, 0){1050}}
}%
{\color[rgb]{0,0,0}\put(1351,-811){\line( 1, 0){300}}
}%
{\color[rgb]{0,0,0}\put(1351,-886){\line( 1, 0){300}}
}%
{\color[rgb]{0,0,0}\put(1351,-961){\line( 1, 0){300}}
}%
{\color[rgb]{0,0,0}\put(1351,-1261){\line( 1, 0){300}}
}%
{\color[rgb]{0,0,0}\put(1351,-1336){\line( 1, 0){300}}
}%
{\color[rgb]{0,0,0}\put(1351,-1411){\line( 1, 0){300}}
}%
{\color[rgb]{0,0,0}\put(2251,-1336){\line( 1, 0){300}}
}%
{\color[rgb]{0,0,0}\put(2251,-1261){\line( 1, 0){300}}
}%
{\color[rgb]{0,0,0}\put(2251,-1411){\line( 1, 0){300}}
}%
{\color[rgb]{0,0,0}\put(3001,-511){\line( 1, 0){300}}
}%
{\color[rgb]{0,0,0}\put(3001,-436){\line( 1, 0){300}}
}%
{\color[rgb]{0,0,0}\put(3001,-586){\line( 1, 0){300}}
}%
{\color[rgb]{0,0,0}\put(3001,-1336){\line( 1, 0){300}}
}%
{\color[rgb]{0,0,0}\put(3001,-1261){\line( 1, 0){300}}
}%
{\color[rgb]{0,0,0}\put(3001,-1411){\line( 1, 0){300}}
}%
{\color[rgb]{0,0,0}\put(2251,-1636){\line( 1, 0){1050}}
}%
{\color[rgb]{0,0,0}\put(2251,-1711){\line( 1, 0){1050}}
}%
{\color[rgb]{0,0,0}\put(2251,-1786){\line( 1, 0){1050}}
}%
{\color[rgb]{0,0,0}\put(3901,-511){\line( 1, 0){300}}
}%
{\color[rgb]{0,0,0}\put(3901,-436){\line( 1, 0){300}}
}%
{\color[rgb]{0,0,0}\put(3901,-586){\line( 1, 0){300}}
}%
{\color[rgb]{0,0,0}\put(3901,-886){\line( 1, 0){300}}
}%
{\color[rgb]{0,0,0}\put(3901,-811){\line( 1, 0){300}}
}%
{\color[rgb]{0,0,0}\put(3901,-961){\line( 1, 0){300}}
}%
{\color[rgb]{0,0,0}\put(3901,-1336){\line( 1, 0){300}}
}%
{\color[rgb]{0,0,0}\put(3901,-1261){\line( 1, 0){300}}
}%
{\color[rgb]{0,0,0}\put(3901,-1411){\line( 1, 0){300}}
}%
{\color[rgb]{0,0,0}\put(3901,-1711){\line( 1, 0){300}}
}%
{\color[rgb]{0,0,0}\put(3901,-1636){\line( 1, 0){300}}
}%
{\color[rgb]{0,0,0}\put(3901,-1786){\line( 1, 0){300}}
}%
{\color[rgb]{0,0,0}\put(1351,-1636){\line( 1, 0){300}}
}%
{\color[rgb]{0,0,0}\put(1351,-1711){\line( 1, 0){300}}
}%
{\color[rgb]{0,0,0}\put(1351,-1786){\line( 1, 0){300}}
}%
{\color[rgb]{0,0,0}\put(2551,-736){\framebox(450,450){$\Phi$}}
}%
{\color[rgb]{0,0,0}\put(2551,-1561){\framebox(450,450){$\Phi$}}
}%
{\color[rgb]{0,0,0}\put(1651,-1861){\framebox(600,1575){\parbox{0.6in}{\centering{swap\\ test}}}}
}%
{\color[rgb]{0,0,0}\put(3301,-1861){\framebox(600,1575){\parbox{0.6in}{\centering{swap\\ test}}}}
}%
{\color[rgb]{0,0,0}\put(2251,-511){\line( 1, 0){300}}
}%
\end{picture}%

  \end{center}
  \caption{\class{QMA} protocol for \prob{Non-isometry}.  The verifier
    accepts only if the first swap test results the symmetric outcome
    and the second swap test results in an antisymmetric outcome.}
  \label{fig:protocol}
\end{figure}

\begin{proto}[\prob{Non-isometry}]\label{proto:non-isometry}
  On an input channel $\Phi \in \transform{H,K}$:
  \begin{enumerate}
  \item Receive a witness state $\rho \in \density{(H \tprod
      R)^{\tprod \text{2}}}$, where $\mathcal{R}$ is a reference space
    such that $\dm{R} = \dm{H}$.
    Apply the swap test to $\rho$,
    rejecting if the outcome is antisymmetric. \label{enum:proto-1}
  \item Use the channel $\Phi$ to obtain 
    $ \sigma = (\Phi \tprod \tidentity{R})^{\tprod 2}(\rho) $. \label{enum:proto-2}
  \item Apply the swap test to $\sigma$, accepting if the outcome is
    symmetric and rejecting otherwise. \label{enum:proto-3}
  \end{enumerate}
\end{proto}
\noindent A diagram of this protocol can be found in
Figure~\ref{fig:protocol}.  The correctness of this protocol is argued in the following theorem.

\begin{theorem}\label{thm:containment}
  Let $\Phi \in \transform{H,K}$, and let $p(\rho)$ be the probability
  that the verifier described in Protocol~\ref{proto:non-isometry} accepts
  the input state $\rho \in \density{(H \tprod R)^{\tprod \text{2}}}$, then
  \begin{enumerate}
  \item If $\min_{\ket\psi} \opnorm{(\Phi \tprod \tidentity{R})(\ketbra{\psi})} \leq
    \varepsilon$, then there exists a witness $\rho$ such that $p(\rho) \geq (1
    - \varepsilon)/2$. \label{enum:proto-proof-1}
  \item If $\min_{\ket\psi} \opnorm{(\Phi \tprod \tidentity{R})(\ketbra{\psi})} \geq
    1 - \varepsilon$, then for any witness $\rho$, $p(\rho) \leq
    9 \varepsilon$.\label{enum:proto-proof-2}
 \end{enumerate}
\end{theorem}
\begin{proof}
    For the sake of brevity, let $\hat{\Phi} = \Phi \tprod
    \tidentity{R}$ throughout.
    To prove the first assertion, let $\ket\psi$ be a pure state in
    $\mathcal{H \tprod R}$ for which $\lVert \hat{\Phi} 
      (\ketbra{\psi}) \rVert_{\infty} \leq \varepsilon$, and let the
    witness state $\rho = \ketbra{\psi} \tprod \ketbra{\psi}$.  This
    state is invariant under the swap operation and so the swap test
    in Step~\ref{enum:proto-1} passes and does not change the state.
    Step~\ref{enum:proto-2} results in the state
    $ [ \hat{\Phi}(\ketbra{\psi}) ]^{\tprod 2} $.  Using
    Equations~\eqref{eqn:purity-largest-eval} and~\eqref{eqn:swaptest-purity},
    the final swap test returns the antisymmetric outcome with
    probability
    \begin{equation*}
      \frac{1}{2} - \frac{1}{2} \tr \left[\hat{\Phi}(\ketbra{\psi})^2 \right]
      \geq \frac{1}{2} - \frac{1}{2} \opnorm{\hat{\Phi}(\ketbra{\psi})}
      \geq \frac{1 - \varepsilon}{2},
    \end{equation*}
    and so the verifier accepts $\rho$ with probability
    approaching one-half for small $\varepsilon$.

    To show the second assertion, we take
    $\hat{\Phi}$ is an $\varepsilon$-isometry and analyze the probability
    that the verifier can be made to accept.  We may assume that the
    witness state lies in the symmetric subspace of $\left( \mathcal{H \tprod R}
    \right)^{\tprod 2}$, as the verifier either rejects in 
    Step~\ref{enum:proto-1} or projects the witness onto this subspace.
    To complete the proof, we show that
     $(\hat{\Phi})^{\tprod 2}$ leaves $\rho$ approximately symmetric.

    To do this, we approximate $\hat{\Phi}$ by an operator that
    preserves the symmetry of input states.  Let $\{ \ket{i} : 1 \leq
    i \leq \dm{H} \}$ be an orthonormal
    basis for the spaces $\mathcal{H}, \mathcal{R}$ (this is possible
    because they have the same dimension).  The states $\{ \ket{ij} : 1 \leq
    i,j \leq \dm{H} \}$ are
    an orthonormal basis for $\mathcal{H \tprod R}$.  Since
    $\hat{\Phi}$ approximately preserves rank, there are states 
    $\ket{\psi_i} \in \mathcal{K}$ such that
    \begin{equation}\label{eqn:defn-psii}
      \tnorm{ (\Phi \tprod \tidentity{R})(\ketbra{ij}) 
        - \ketbra{\psi_i} \tprod \ketbra{j} } \leq \varepsilon
    \end{equation}
    for all $i$ and $j$.  We define a linear operator $A \colon
    \mathcal{H} \to \mathcal{K}$ by the equation $A \ket{i} =
    c_{i} \ket{\psi_i}$, where the $c_i \in \mathbb{C}$ with
    $\abs{c_i} = 1$.  The introduction of the phases $c_i$ is necessary because
    Equation~\eqref{eqn:defn-psii} only defines the states
    $\ket{\psi_i}$ up to a phase.
    Note that the operator $A$ is not necessarily
    unitary as we may not assume that the states $\ket{\psi_i}$
    are orthogonal.  The next step is to show that, for some choice of the phases
    $c_i$, conjugation by $A$ approximates the channel $\Phi$ in the
    trace norm.  This is the most technical portion of the proof.
    
    Consider the output of $\hat{\Phi}$ on the entangled
    state $(\ket{ii} + \ket{jj})/\sqrt{2}$ in $\mathcal{H \tprod R}$,
    given by
   \begin{equation}\label{eqn:defn-rho}
      \rho 
     = \frac{1}{2} \sum_{a,b \in \{i,j\}} 
      \hat{\Phi}(\ket{aa}\bra{bb})
      = \frac{1}{2} \sum_{a,b \in \{i,j\}} 
      \Phi(\ket{a}\bra{b}) \tprod \ket{a} \bra{b}.
    \end{equation}
    Since $\hat{\Phi}$ maps pure states to states that are nearly
    pure, we know that the purity of $\rho$ satisfies $\tr(\rho^2)
    \geq (1 - \varepsilon)^2 \geq 1 - 2 \varepsilon$.  Evaluating the
    purity using Equation~\eqref{eqn:defn-rho} gives
    \begin{align}
      1 - 2 \varepsilon
      \leq \tr(\rho^2)
       &= \frac{1}{4} \left( \tr \Phi(\ketbra{i})^2 + \tr \Phi(\ketbra{j})^2
         + 2 \tr \Phi( \ket i \bra j ) \Phi( \ket j \bra i ) \right)
       \nonumber \\
      &\leq \frac{1}{2} 
        + \frac{1}{2} \tr \left(( \Phi(\ket i \bra j) \Phi(\ket i \bra
          j)^* \right).
        \label{eqn:sval-bound}
    \end{align}
    Interpreting the expression $\tr X X^*$ as the sum of the squared
    singular values of $X$, Equation~\eqref{eqn:sval-bound}
    implies that the operator $\Phi(\ket i \bra j)$ has largest
    singular value at least $1 - 4 \varepsilon$.  
    Since the sum of the singular values of this
    operator cannot exceed one (as the trace norm does not increase
    under the application of a channel), this implies that it can be
    decomposed as
    \begin{equation}\label{eqn:defn-cd}
      \Phi(\ket i \bra j) = (1 - 4 \varepsilon) \ket{\phi_i} \bra{\phi_j} + 4 \varepsilon Y,
    \end{equation}
    where $\ket{\phi_i}, \ket{\phi_j} \in \mathcal{K}$ are pure and $Y$ is
    a linear operator on $\mathcal{K}$ with $\tnorm{Y} = 1$.
    It remains to show that the vectors $\ket{\phi_i}$ and
    $\ket{\phi_j}$ are, up to a phase, approximately equal to the vectors
    $\ket{\psi_i}$ and $\ket{\psi_j}$ defined in
    Equation~\eqref{eqn:defn-psii}.  To do this, we consider the
    action of $\Phi$ on $(\ket{i} + \ket{j})/\sqrt{2}$.  Since
    $\Phi$ is an $\varepsilon$-isometry, the
    output of $\Phi$ on this state is within trace distance $2
    \varepsilon$ of some pure state $\ket{\gamma}$.  Combining
    Equations~\eqref{eqn:defn-psii} and~\eqref{eqn:defn-cd} and
    applying the triangle inequality yields
    \begin{equation*}
      \tnorm{\ketbra \gamma - 
        \frac{1}{2} \left(\ketbra{\psi_i} + \ket{\phi_i} \bra{\phi_j}  + 
        \ket{\phi_j} \bra{\phi_i} + \ketbra{\psi_j} \right)} \leq 5 \varepsilon.
    \end{equation*}
    Since $\ket\gamma$ is pure, for some phases $c_i$
    and $c_j$ we have
    $ \lVert \ket{\phi_i} \bra{\phi_j} - c_i c_j^* \ket{\psi_i}
        \bra{\psi_j}  \rVert_{\mathrm{tr}} 
      \leq 5 \varepsilon$,
%     \begin{equation*}
%       \tnorm{ \ket{\phi_i} \bra{\phi_j} - c_i c_j^* \ket{\psi_i}
%         \bra{\psi_j} } 
%       \leq 5 \varepsilon,
%     \end{equation*}
    which in turn implies that
    $  \lVert \Phi(\ket i \bra j) - c_i c_j^* \ket{\psi_i}
        \bra{\psi_j} \rVert_{\mathrm{tr}} \leq 9 \varepsilon$,
%     \begin{equation*}
%       \tnorm{ \Phi(\ket i \bra j) - c_i c_j^* \ket{\psi_i}
%         \bra{\psi_j} } \leq 9 \varepsilon,
%     \end{equation*}
    using Equation~\eqref{eqn:defn-cd}.  Finally, since this is true
    for any $i \neq j$, and the case of $i = j$ is
    Equation~\eqref{eqn:defn-psii}, the previous equation implies that    
    $\max_{\rho} \tnorm{\Phi(\rho) - A \rho A^*} \leq 9 \varepsilon$,
%     \begin{equation}\label{eqn:phi-approx-a}
%       \max_{\rho} \tnorm{\Phi(\rho) - A \rho A^*} \leq 9 \varepsilon,
%     \end{equation}
    where $A$ is the operator defined by $A \ket{i} = c_i \ket{\psi_i}$ for
    all $i$.
    
    It remains only to show that the operator $A \tprod A$ preserves
    symmetric states.  To see this, take $\ket{ij} + \ket{ji}$ an
    arbitrary basis element of the symmetric subspace of
    $\mathcal{H}^{\tprod 2}$.  By a simple calculation
    \begin{equation*}
      (A \tprod A)(\ket{ij} + \ket{ji})
      = c_i c_j \ket{\psi_i} \tprod \ket{\psi_j} 
      + c_i c_j \ket{\psi_j} \tprod \ket{\psi_i},
    \end{equation*}
    which remains invariant under swapping the two spaces.  By
    linearity, conjugation by $A \tprod \identity{R}$
    also preserves the symmetry of states on $(\mathcal{H \tprod
      R})^{\tprod 2}$.  It
    follows that $\hat{\Phi}$ preserves symmetry
    up to an error of $9 \varepsilon$ in the trace distance.  This
    implies that the swap test on the output of $\hat{\Phi} \tprod
    \hat{\Phi}$ applied to a symmetric state returns the symmetric
    outcome with probability at least $1 - 9 \varepsilon$. \qed 
\end{proof}

\noindent This theorem shows that $\prob{Non-isometry}_\varepsilon$
is in \class{QMA} for any constant $\varepsilon$ satisfying $(1 -
\varepsilon)/2 > 9 \varepsilon$.  Together with the
\class{QMA}-hardness of the problem shown in
Theorem~\ref{thm:hardness} this gives the main result.

\begin{corollary}\label{cor:main-result}
  For any constant $\varepsilon < 1 / 19$,
  $\prob{Non-isometry}_\varepsilon$ is \class{QMA}-complete.
\end{corollary}

\noindent This also implies that problem of computing the channel purity, as defined by
Zanardi and Lidar~\cite{ZanardiL04}, over the whole input space is \class{QMA}-complete.

\section{Conclusion}

We have shown the computational intractability of the problem of detecting when a
quantum channel is far from an isometry, or equivalently, when a
channel can be made to output a highly mixed state.  These results
show that it is extremely difficult to characterize the worst-case
behaviour of a quantum computation.  This is similar to the classical
case, where the problem of determining if a circuit can produce a
specific output is known to be intractable.

We have also added to the short but growing list of problems that are
known to be complete for the complexity class \class{QMA}.  The
\prob{Non-isometry} problem provides a new way to study this class, as
it exactly characterizes the difficulty of the problems in the class.
It is hoped that this will lead to new results about the power of this
model of computation.

% There are several open problems related to this work.  A few of the
% more interesting ones are listed below.
% \begin{itemize}
%   \item As the verifier in a \class{QMA} protocol may use ancillary
%     qubits, the hardness proof in Theorem~\ref{thm:hardness} only
%     applies to isometries.  If a protocol could be constructed without
%     these ancillary qubits, the same argument would apply to the case
%     of testing unitarity.

%   \item The bound of $9 \varepsilon$ in Item~\ref{enum:proto-proof-2}
%     of Theorem~\ref{thm:containment} is hardly expected to be
%     optimal.  Improving this argument would put the problem into
%     \class{QMA} for larger values of $\varepsilon$ than $1/19$.

%   \item If we take an alternate definition of approximate isometry
%     which is that the Choi matrix is close to a maximally entangled we
%     end up with a weaker notion of isometry.  This notion can be
%     tested in \class{BQP} using the swap test to estimate the purity
%     of the Choi matrix, as this definition avoids the minimization of
%     purity over all input states.  Is this simpler problem complete
%     for \class{BQP}?
% \end{itemize}

\section*{Acknowledgements}

I am grateful for discussions with Markus Grassl, Masahito Hayashi,
Lana Sheridan, and John Watrous, from which I have learnt a great
deal.
This work has been supported by the Centre for Quantum Technologies,
which is funded by the Singapore Ministry of Education and the
Singapore National Research Foundation, as well as as well the Bell
Family Fund, while the author was at the Institute for Quantum
Computing at the University of Waterloo.

%\bibliographystyle{abbrv-modern}
%\bibliography{mathphys}

\begin{thebibliography}{10}
 \providecommand{\doi}[1]{{\sc doi}: \href{http://dx.doi.org/#1}{#1}}
 \providecommand{\urlprefix}{{\sc url} }
 \providecommand{\eprintprefix}{{\sc eprint}: }

\bibitem{AharonovK+98}
D.~Aharonov, A.~Kitaev, and N.~Nisan.
\newblock Quantum circuits with mixed states.
\newblock In \emph{Proceedings of the 30th ACM Symposium on the Theory of
  Computing}, pp. 20--30. 1998.
\newblock \doi{10.1145/276698.276708}.
\newblock \eprintprefix\oldarxiv{9806029}.

\bibitem{BeigiS07}
S.~Beigi and P.~W. Shor.
\newblock On the complexity of computing zero-error and {Holevo} capacity of
  quantum channels, 2007.
\newblock \eprintprefix\arxiv{0709.2090v3}.

\bibitem{BuhrmanC+01}
H.~Buhrman, R.~Cleve, J.~Watrous, and R.~de~Wolf.
\newblock Quantum fingerprinting.
\newblock \emph{Physical Review Letters}, 87(16):167902, 2001.
\newblock \doi{10.1103/PhysRevLett.87.167902}.
\newblock \eprintprefix\oldarxiv{0102001}.

\bibitem{Choi75}
M.-D. Choi.
\newblock Completely positive linear maps on complex matrices.
\newblock \emph{Linear Algebra and its Applications}, 10(3):285--290, 1975.
\newblock \doi{10.1016/0024-3795(75)90075-0}.

\bibitem{EkertA+02}
A.~K. Ekert, C.~M. Alves, D.~K. Oi, M.~Horodecki, P.~Horodecki, and L.~C. Kwek.
\newblock Direct estimations of linear and nonlinear functionals of a quantum
  state.
\newblock \emph{Physical Review Letters}, 88(21):217901, 2002.
\newblock \doi{10.1103/PhysRevLett.88.217901}.
\newblock \eprintprefix\oldarxiv{0203016}.

\bibitem{JanzingW+05}
D.~Janzing, P.~Wocjan, and T.~Beth.
\newblock ``{Non-identity-check}'' is {QMA}-complete.
\newblock \emph{International Journal of Quantum Information}, 3(3):463--473,
  2005.
\newblock \doi{10.1142/S0219749905001067}.
\newblock \eprintprefix\oldarxiv{0305050}.

\bibitem{JiW09}
Z.~Ji and X.~Wu.
\newblock Non-identity check remains {QMA}-complete for short circuits, 2009.
\newblock \eprintprefix\arxiv{0906.5416}.

\bibitem{KempeK+06}
J.~Kempe, A.~Kitaev, and O.~Regev.
\newblock The complexity of the local {Hamiltonian} problem.
\newblock \emph{{SIAM} Journal on Computing}, 35(5):1070--1097, 2006.
\newblock \doi{10.1137/S0097539704445226}.
\newblock \eprintprefix\oldarxiv{0406180}.

\bibitem{Kitaev99}
A.~Y. Kitaev.
\newblock Quantum {NP}.
\newblock Talk at the 2nd Workshop on Algorithms in Quantum Information
  Processing (AQIP), DePaul University, 1999.

\bibitem{KitaevS+02}
A.~Y. Kitaev, A.~H. Shen, and M.~N. Vyalyi.
\newblock \emph{Classical and Quantum Computation}, volume~47 of \emph{Graduate
  Studies in Mathematics}.
\newblock American Mathematical Society, 2002.

\bibitem{Knill96}
E.~Knill.
\newblock Quantum randomness and nondeterminism.
\newblock Techical Report LAUR-96-2186, Los Alamos National Laboratory, 1996.
\newblock \eprintprefix\oldarxiv{9610012}.

\bibitem{Liu06}
Y.-K. Liu.
\newblock Consistency of local density matrices is {QMA}-complete.
\newblock In \emph{Proceedings of the 10th International Workshop on Randomized
  Techniques in Computation}, volume 4110 of \emph{Lecture Notes in Computer
  Science}, pp. 438--449. Springer, 2006.
\newblock \doi{10.1007/11830924\_40}.
\newblock \eprintprefix\oldarxiv{0604166}.

\bibitem{LiuC+07}
Y.-K. Liu, M.~Christandl, and F.~Verstraete.
\newblock Quantum computational complexity of the {N}-representability problem:
  {QMA} complete.
\newblock \emph{Physical Review Letters}, 98(11):110503, 2007.
\newblock \doi{10.1103/PhysRevLett.98.110503}.
\newblock \eprintprefix\oldarxiv{0609125}.

\bibitem{MarriottW05}
C.~Marriott and J.~Watrous.
\newblock Quantum {Arthur-Merlin} games.
\newblock \emph{Computational Complexity}, 14(2):122--152, 2005.
\newblock \doi{10.1007/s00037-005-0194-x}.
\newblock \eprintprefix\oldarxiv[cs]{0506068}.

\bibitem{NielsenC00}
M.~A. Nielsen and I.~L. Chuang.
\newblock \emph{Quantum Computation and Quantum Information}.
\newblock Cambridge University Press, 2000.

\bibitem{SchuchC+08}
N.~Schuch, I.~Cirac, and F.~Verstraete.
\newblock Computational difficulty of finding matrix product ground states.
\newblock \emph{Physical Review Letters}, 100(25):250501, 2008.
\newblock \doi{10.1103/PhysRevLett.100.250501}.
\newblock \eprintprefix\arxiv{0802.3351}.

\bibitem{SchuchV09}
N.~Schuch and F.~Verstraete.
\newblock Computational complexity of interacting electrons and fundamental
  limitations of density functional theory.
\newblock \emph{Nature Physics}, 5(10):732 -- 735, 2009.
\newblock \doi{doi:10.1038/nphys1370}.
\newblock \eprintprefix\arxiv{0712.0483}.

\bibitem{Watrous00}
J.~Watrous.
\newblock Succinct quantum proofs for properties of finite groups.
\newblock \emph{Proceedings of the 41st {IEEE} Symposium on Foundations of
  Computer Science}, pp. 537 -- 546, 2000.
\newblock \doi{10.1109/SFCS.2000.892141}.
\newblock \eprintprefix\oldarxiv[cs]{0009002}.

\bibitem{WeiM+10}
T.-C. Wei, M.~Mosca, and A.~Nayak.
\newblock Interacting boson problems can be {QMA} hard.
\newblock \emph{Physical Review Letters}, 104(4):040501, 2010.
\newblock \doi{10.1103/PhysRevLett.104.040501}.
\newblock \eprintprefix\arxiv{0905.3413}.

\bibitem{ZanardiL04}
P.~Zanardi and D.~A. Lidar.
\newblock Purity and state fidelity of quantum channels.
\newblock \emph{Physical Review A}, 70(1):012315, 2004.
\newblock \doi{10.1103/PhysRevA.70.012315}.
\newblock \eprintprefix\oldarxiv{0403074}.

\end{thebibliography}
\newcommand{\arxiv}[2][quant-ph]{\href{http://arxiv.org/abs/#2}{arXiv:#2 [#1]}}
  \newcommand{\oldarxiv}[2][quant-ph]{\href{http://arxiv.org/abs/#1/#2}{arXiv:%
#1/#2}}

\end{document}